# *Excitonic Insulator and Possible Superfluid Based on Two-Dimensional Diamond*

*Shisheng Lin[1,2]\*, Shaoqi Huang[1], Minhui Yang[1], Xin Chen[1], Hongjia Bi[1], Kangchen Xiong[1]*

[1] College of Information Science and Electronic Engineering, Zhejiang University; Hangzhou, 310027, China.

2 State Key Laboratory of Extreme Photonics and Instrumentation, Zhejiang University; Hangzhou, 310027, China.

\*Corresponding author. Email: shishenglin@zju.edu.cn

## Abstract

Recent research on excitonic insulator has progressed mainly based on narrow bandgap semiconductor or semimetal. Herein, we realize excitonic insulator based on two-dimensional (2D) wide band gap diamond with transition temperature as high as 220K. The resistance rises dramatically by more than three orders, which can be explained by the Bose-Einstein condensation (BEC) of excitons. While cooling down below transition temperature, the wavelength of the bound excitons caused by boron and nitrogen centers becomes highly overlapped, leading to BEC process. Furthermore, the variable range hopping mechanism is used to simulate the resistance as a function of temperature, which reveals the formation of excitonic insulator. When temperature drops down further, a sudden drop of resistance over three orders was observed around 60K, possibly due to the formation of non-equilibrium excitonic superfluid resulting from highly overlap of wavelength of the large density bound excitons at lower temperature. This study provides evidences for excitonic insulator and possible superfluid phase based on wide bandgap semiconductor.

# Introduction

Excitonic insulator (EI) is a crucial state of matter, the gap of which is related to the binding energy of excitons and their collective ground state, rather than the band structure of the material itself[1]. Unlike conventional conductors or insulators where electrons and holes move independently, excitons in EI behave as composite bosons, undergoing condensation into a macroscopically coherent state, especially in two-dimensional (2D) system.[2-6] The condensed excitons can exhibit quantum coherence over macroscopic distance, similar to the coherence observed in superfluid helium and Bose-Einstein condensation (BEC), offering promising pathways for applications such as polariton lasers and quantum computing[7,8]. Critically, EI can serve as a parent material of exciton-mediated superconductor. Through strong exciton-electron interactions, doped electrons can be transformed into Cooper pairs, potentially enabling high-temperature superconductivity as theorized by many scientists including John Bardeen[9-14]. However, the search for materials that can sustain an EI phase is challenging, requiring systems with a strong interaction between excitons.

Compared with bulk materials, low-dimensional system can exhibit significant enhanced exciton binding energy ($E_b$) due to quantum confinement effect, which suppresses dielectric screening and amplifies Coulomb interactions, providing larger possibility to realize EI[15,16]. Normally, heterostructures made from layering 2D materials can exhibit strong Coulomb coupling while spatially separating electrons and holes to form stable excitons. Layered InAs/GaSb has been adapted to explore EI as $E_b$ is larger than the band gap of either InAs or GaSb[17,18]. Recent advances also highlight $Ta_2NiSe_5$ as a prime EI candidate, exhibiting a transition temperature as high as 325K[19]. However, most of the abovementioned researches mainly focused on narrow bandgap materials or semimetals, achieving EI phase by opening or expanding an external gap introduced by excitons[20,21]. Wide band gap materials like diamond remain underexplored despite their potential for EI phase with high transition temperature. When confined to 2D thin film, diamond exhibits $E_b$ as large as several hundred meV, far exceeding thermal energy at 300K (26meV), so it is possible to host robust EI phase.[22,23]

Herein, by co-doping boron acceptors with nitrogen donors, we realized EI in 2D single crystal diamond with thickness below 15nm. The resistance can rise dramatically by more than three orders with a transition temperature ($T_C$) of 220K. We successfully described the transition and insulating phase with variable range hopping (VRH) model, reasonably explaining the transport behavior by the BEC of excitons below $T_C$. When temperature crosses 61.1K, the resistance quickly drops from 12 MΩ at 61.1 K down to 2.5kΩ at 23.1K, possibly marking the second phase transition from EI to the non-equilibrium superfluid of bound excitons, and the resistance behaves as an inverse proportional function with the biased current at temperature below 20K. This research provides the evidence for EI phase and possible superfluid phase in wide band gap system, which could open the avenue of high temperature superconductor or superfluid through exciton designment.

## Result And Discussion

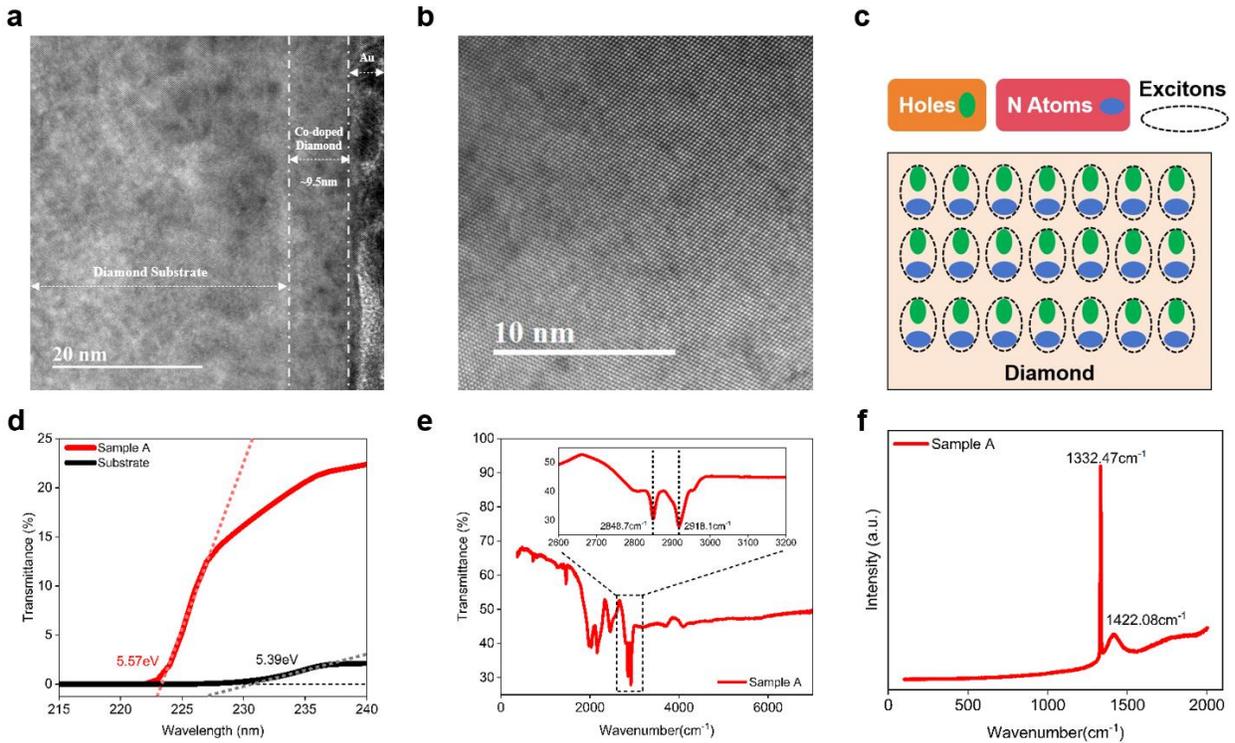

**Fig.1 | Characterization of 2D diamond. a** HRTEM image of the cross section of 2D diamond and diamond substrate. **b** Atomic structure of diamond along <001> direction of the sample. **c** Schematic illustration of bound excitons in B-N co-doped diamond. **d** Ultraviolet-visible (UV-Vis) spectroscopy of sample A and pure diamond substrate. **e** Infrared (IR) spectroscopy of the sample. **f** Raman spectrum of the sample.

Fig. 1a shows a high-resolution transmission electron microscopy (HRTEM) image of a typical boron-nitrogen-co-doped diamond sample. The film was grown homoepitaxially on a single crystal diamond substrate, with a thickness of approximately 9.5nm, as marked in the figure. Four gold electrodes were sputtered in parallel on the 6mm×6mm surface for four-probe measurements. Fig. 1b shows the HRTEM image of the diamond along <001> direction, revealing a well-defined atomic lattice that confirms high crystal quality of the diamond film, laying the foundation for BEC of excitons. Another repeating sample with its HRTEM image and FFT result is shown in Supplementary Fig. 1. Fig. 1c schematically illustrates nitrogen deep-donor centers in diamond, where holes are localized or shared by the nitrogen centers, forming the bound excitons. Since the reduced dielectric screening in 2D system will enhance the probability of excitonic ground state formation, fabricating 2D diamond will facilitate EI phase by leveraging these bound states. Herein, 2D indicates that boron or nitrogen centers locate at the surface of epitaxial ultrathin diamond, which means bound excitons can only appear at the epitaxial 2D diamond. Fig. 1d compares ultraviolet-visible (UV-Vis) spectroscopy of sample A and its single crystal diamond substrate. Through linear extrapolation, the band gap of the diamond substrate can be derived as 5.39eV, indicating the purity of single crystal diamond. The blue-shifted absorption edge at 5.57 eV in the co-doped sample arises from 2D confinement effect, which increases the band gap of the epitaxial diamond. We should emphasize that the BEC of bound excitons does not require that $E_b$ should larger than the band gap especially in wide band gap semiconductor, although it is required in narrow band gap semiconductors such as $Ta_2NiSe_5$[24]. Another UV-Vis spectroscopy result is shown in Supplementary Fig. 2 for sample B2, where the absorption edge increases to about 5.54eV indicating an enlarged band gap by 2D confinement effect, similar with sample A. From Fig. 1e, two peaks at 2848.7cm$^{-1}$ and 2918.1cm$^{-1}$ in the infrared spectroscopy of the sample are clearly depicted, representing deep and shallow doped boron acceptors respectively[25]. Fig. 1f shows the Raman spectrum of diamond, where the 1332.47cm$^{-1}$ peak is ascribed to the first-order Raman scattering caused by the zero-center optical phonon mode, indicating no stress has been introduced into the epitaxial grown diamond. The peak at 1422.08cm$^{-1}$ is resulting from N$_V$ color centers, confirming the heavy doping of nitrogen[26]. Collectively, these data and figures demonstrate successful growth of epitaxial high quality single crystal 2D diamond, laying the groundwork for the EI phase based on bound excitons.

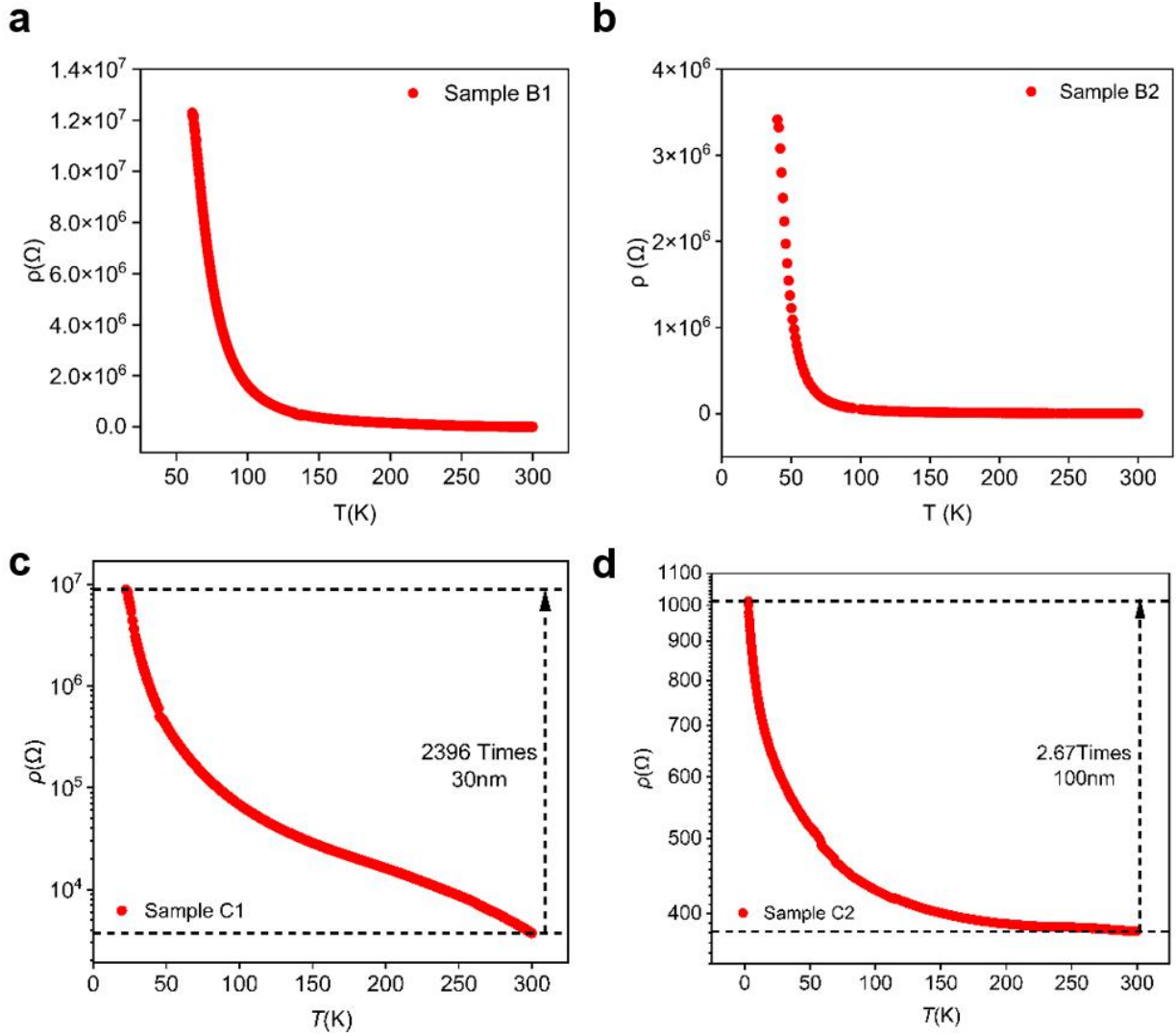

**Fig.2 | Transport characteristic of EI based on 2D diamond. a** Sample B1 and **b** sample B2 exhibiting EI phase, where resistance increases rapidly as the temperature decreases. **c-d** Comparative transport behavior of sample C1 and C2, whose thickness are 30nm and 90nm respectively.

Fig. 2a shows the resistance as a function of temperature for sample B1 with a thickness of 9nm. The resistance exhibits a dramatic increase from 6 kΩ at 300 K to nearly 12 MΩ at 61.1 K as temperature decreases. Basically, lower temperature suppresses the thermal ionization of boron acceptors, reducing the concentration of free charge carriers. On the other hand, the wavelength of stable excitons formed at low temperature by boron and nitrogen centers becomes larger, causing the strong interaction among the exciton bosons and finally the formation of BEC. The free charge carriers are furtherly confined by the excitons through Coulomb interaction, reducing the conductivity and finally leading to the EI phase. Sample B2 in Fig. 2b shows similar resistance behavior with sample B1, validating the reproducibility of the EI phase of 2D diamond. Other three samples (B3-B5) measured with repeatable functions of resistance versus temperature in different instruments are shown in Supplementary Fig. 3. The thickness of all the samples mentioned is below 15nm, and the resistance increases more than three orders while temperature cools down for most of the samples. It should be pointed out that 2D confinement effect makes great contribution to the formation of EI phase in diamond. For comparison, we fabricated three additional samples labeled as C1, C2 and C3 with thickness of 30nm, 90nm, and more than 1000nm, respectively. Fig. 2c shows the resistance of sample C1 as a function of temperature, which still retains EI behavior with a resistance increase by a factor of 2396 from 300 K to 22 K. However, as sample thickness further increases (Fig. 2d, sample C2, 90nm), the rate of increment of resistance significantly diminishes compared with 2D system. In heavily doped bulk sample C3 with a thickness of more than 1000nm shown in Supplementary Fig. 4, the resistance does not increase continuously as temperature decreases, and it varies by less than 20% from 300K to 3K. The abovementioned comparative results indicate that with the increment of the thickness of the samples, the B-N co-doped diamond gradually lose the behavior of an EI. Therefore, the increasement of resistance in 2D diamond as a function of deceased temperature should be attributed to EI phase formed by bound excitons confined in 2D space.

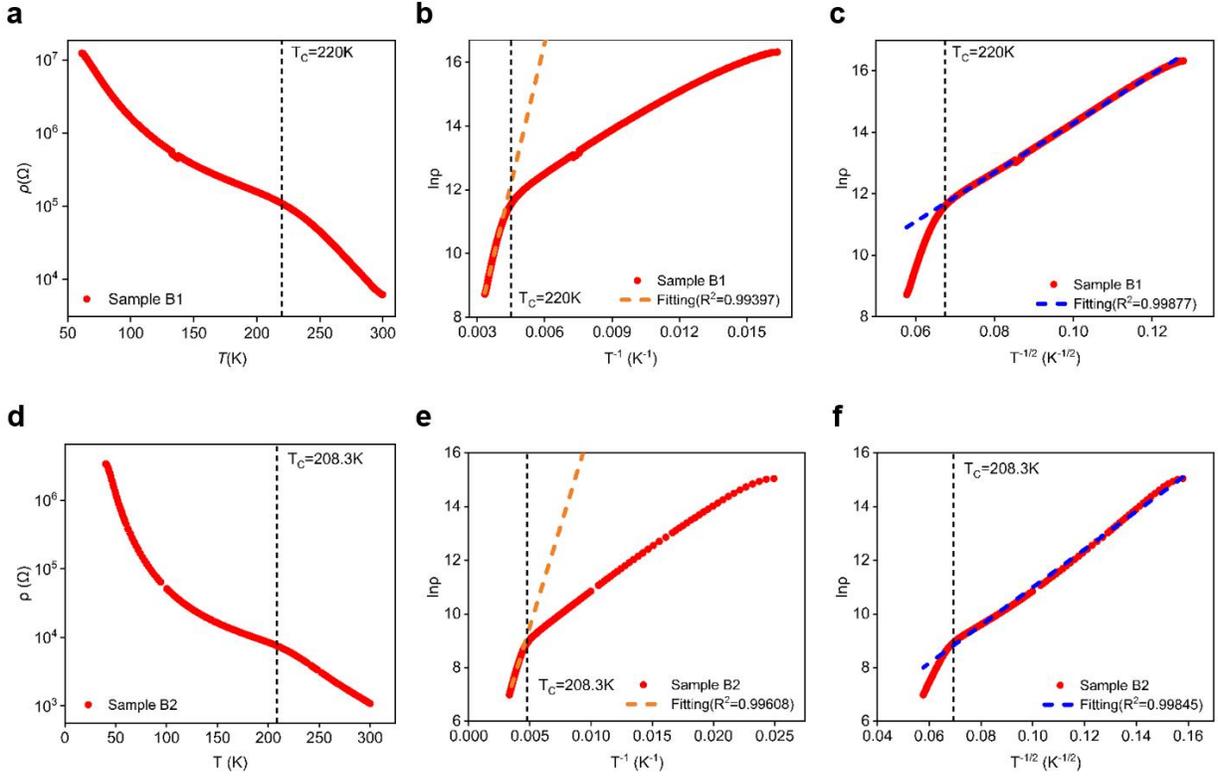

**Fig.3 | Resistance model of EI based on 2D diamond. a** Resistance as a function of temperature in logarithmic form for sample B1. Transition temperature $T_C$ = 220K is marked on the figure. **b** Change of resistance fits well with Arrhenius model for temperature above $T_C$. **c** Change of resistance fits well with variable range hopping (VRH) model for temperature below $T_C$, where excitons condense into ground state. **d-f** Resistance in logarithmic form, fitting of Arrhenius model for temperature above $T_C$ and fitting of VRH model for temperature below $T_C$ for sample B2, respectively. Transition temperature $T_C$ = 208.3K is marked on each figure. Coefficient of determination ($R^2$) for each fitting curve is closed to 1.0, indicating the good linearity.

We selected sample B1 and B2 for in-depth analysis. Fig. 3a shows the logarithmic temperature-dependent resistance, clearly revealing an increase from less than 10kΩ in room temperature to more than 10MΩ at 61.1K. A slight but clear change in the slope of resistance occurs around 220K, indicating the transition from normal phase into EI phase, below which the transport behavior of the sample is dominated by exciton-mediated localization due to excitons condensation, instead of thermal activation. At temperature above $T_C$, the transport characteristic fits pretty well with Arrhenius model[27] given by equation (1), as shown in Fig. 3b.

$$\rho = \rho_0 \exp\left(\frac{E_A}{k_B T}\right) \tag{1}$$

Where $\rho$ is the resistance of the sample, $\rho_0$ is a coefficient unrelated to temperature, $E_A$ is the activation energy, characterizing the difficulty for electrons to be thermally excited, $k_B$ is Boltzmann's constant and $T$ stands for temperature. The $ln\rho$-$T^{-1}$ function in Fig. 3b shows great linearity above 220K with a coefficient of determination ($R^2$) closed to 1.0, confirming a complete thermal excitation ($E_A \approx 0.254$eV) dominated by ionized boron acceptors. When temperature continuously decreases, the $ln\rho$-$T^{-1}$ curve of sample B1 steepens significantly within a small range of temperature around 220K, which is marking the phase transition of the excitons in 2D diamond. Below 220K, the wavelength of the excitons becomes larger and overlapped, so excitons are condensed together, strongly suppressing the flow of free charge carriers, leading to the EI phase. In the low temperature area, since the carriers are localized by bound excitons composed of randomly distributed boron and nitrogen atoms, they tend to hop among localized bound exciton centers, which is known as variable range hopping (VRH) model predicted by N. F. Mott[28,29]. In an interacting system like B-N co-doped 2D diamond, VRH model indicates that the resistance will follow Efros and Shklovskii's Law[30], given by equation (2):

$$\sigma(T) \propto \exp\left[-\left(\frac{T_0}{T}\right)^{\frac{1}{2}}\right] \tag{2}$$

Where $T_0$ is the characteristic temperature given by equation (3)[31]:

$$T_0 = \frac{\beta e^2}{\varepsilon k_B \xi} \tag{3}$$

$\beta$=6.5 for 2D system[32], $e$ stands for elementary charge, $\varepsilon$ is the dielectric constant for material and $\xi$ represents the localization length, which defines the spatial extent of the wavefunction of electron. From Fig. 3c, it is seen that the experimental resistance of the sample fits well with VRH mathematical model below $T_C$ with a large $R^2$ close to 1.0, confirming the behavior of variable range hopping of carriers, thus implying the stable presence of EI phase for B-N co-doped 2D diamond. It is calculated that the localization length $\xi \approx 38.9$nm, revealing a large spatial extent of electron wavefunction. Additionally, since the localization length of electrons is much larger than the Bohr radius of excitons in diamond (<1nm)[33,34], the excess holes contributed by boron acceptors can strengthen the strongly interaction among excitons in a large scale. Fig. 3d-f shows another sample B2 for repeated confirmation, with a resistance from

1080Ω at room temperature to more than 3.4 kΩ around 42K. The marked transition temperature $T_C$ is about 208.3K, and the localization length $\xi$ is 49.8nm. Another two samples (B3, B4) and their fitting models are shown in supplementary Fig. 5, both of which exhibit similar dependence of resistance on temperature like sample B1 and B2, further confirming the reproducible EI phase based on 2D co-doped diamond. As analyzed above, the localization length $\xi$ is inferred to be around 40 nm, so diamond film with thickness less than 40 nm such as sample C1, should have similar EI behavior at low temperature, which agrees well with the experimental results presented in Fig. 2c.

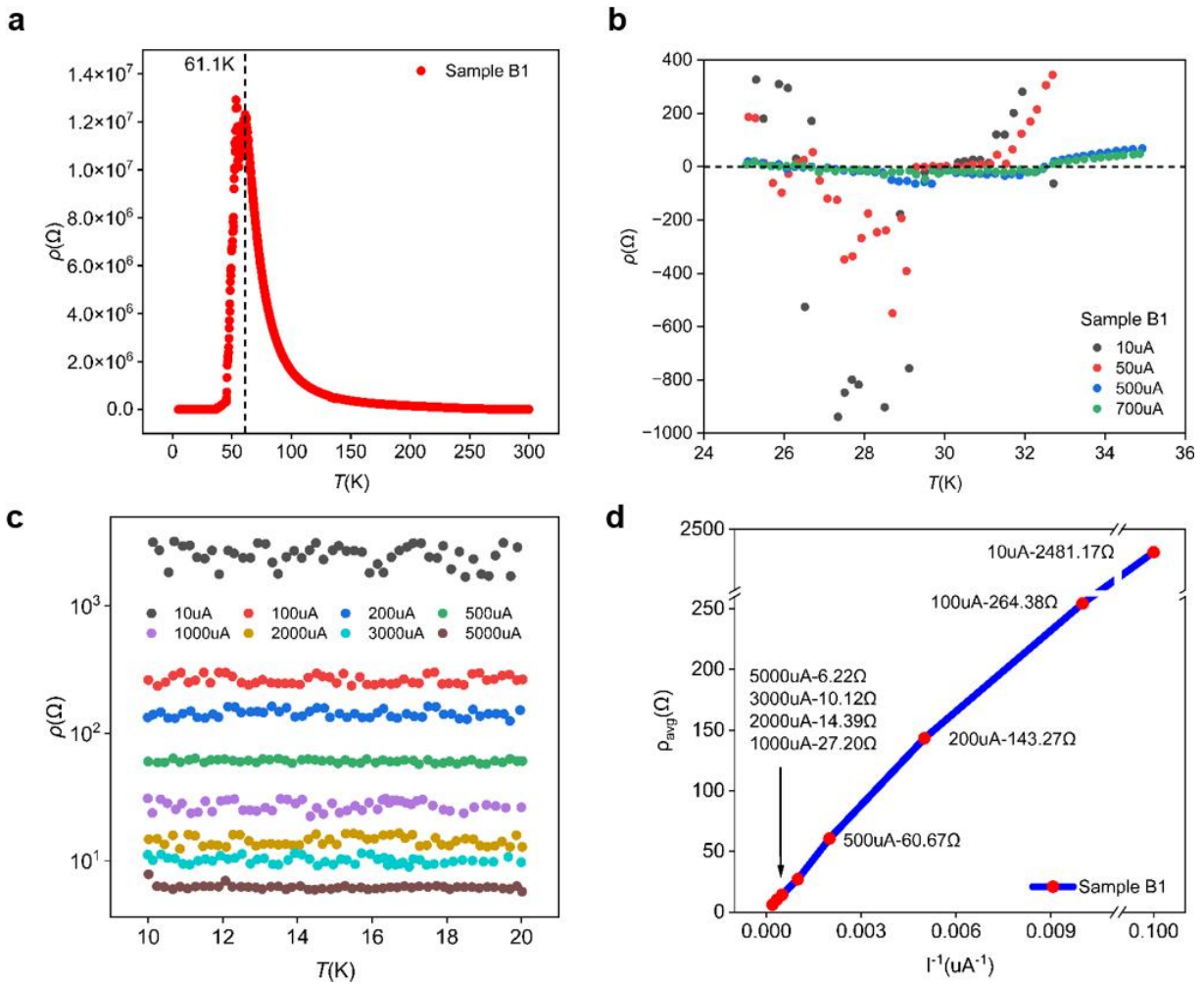

**Fig.4 | Possible excitonic mediated superfluidity. a** Sharp drop of resistance for sample B1 at 61.1K. **b** Resistance crosses over zero near 30 K for applied current of 10μA, 50μA, 500μA and 700μA. **c** Dependence of resistance as a function of temperature below 20K, where the resistance

is inverse proportion to the applied current. **d** Dependence of resistance as a function of the biased current, where an inverse linear function can be clearly clarified.

It should be emphasized that the resistance of the sample B1 exhibit an abrupt drop to 2.5 kΩ after reaching the highest point around 61.1K, significantly marking the second phase transition of 2D co-doped diamond, as shown in Fig. 4a. Moreover, negative resistance occurs around 30K as shown in Fig. 4b and the resistance remains stable relatively at extremely low temperature below 20K. As shown in Fig. 4c-d, by increasing the flowing current through the sample at the value of 10μA, 100μA, 200μA, 500μA, 1000μA, 2000μA, 3000μA and 5000μA, respectively, the average resistance decreases linearly as a function of flowing current.

We tend to recognize these striking transport anomalies as the transition from EI phase into superfluid phase of the bound excitons. The drop of resistance could be induced by the formation of non-equilibrium superfluid, as the holes contributed by boron are much more than localized nitrogen centers. As temperature lowers, the wavelength of bound excitons becomes larger, so the Coulomb interaction among the excitons becomes stronger, which leads to the possible superfluid behavior. At the critical temperature, bound excitons form due to strong Coulomb attraction which overcomes screening. The system gaps out at the Fermi surface, similar to how Cooper pairing gaps out a metal in superconductivity. As temperature decreases below the EI transition, thermal fluctuations are suppressed. The phase coherence between excitons can be strengthened, and if the excitons not only exist as bound objects but also gain long-range phase coherence, their collective motion can emerge. This would manifest as dissipationless counterflow transport, where electrons flow in one direction and holes in the opposite, while the relative motion of the two species is frictionless[35,36]. Once the EI forms, lowering temperature further can indeed enhance coherence and lead to a superfluid state of excitons, characterized by frictionless exciton flow, which has been reproducibly found in Fig. 4 and Supplementary Fig. 6. As seen from Fig. 4, the possible exciton superfluid exhibits some resistance, meaning it has some friction among excitons during the movement, which should be induced by non-equilibrium status of EI. Actually, in Fig. 4, when temperature lowers down to 20K, the superfluid phase remains stable and the resistance stay fluctuating. As the exerted current increases, the relationship between resistance and biased current follows the inverse proportional function as shown in Fig. 4c-d. As the interaction strength between holes contributed by boron and bound excitons increases, the chance of formation of Cooper pairs increase, which leads to the decreased resistance[37]. The negative resistance is possibly because some of the electrons or holes are pulled out by the strong interaction between the Cooper pairs and excitons, resulting in a reversed current direction in the method of four-probe resistance measurement. Another sample possibly behaves as a superfluid named B5 is presented in Supplementary Fig. 6, where similar transport characteristics occur again through measured on different instruments.

We emphasize that this is a novel area for EI in 2D large band gap semiconductor. More experiments and theoretical calculations should be carried out in the following works, which requires the efforts of the whole scientific society on this area. However, it is repeatedly confirmed that a transition from EI phase to a second phase occurs in 2D diamond, confirmed by an abrupt drop of resistance and the consistence between experimental results and simulated results. Delicate experiments are needed to be designed in the future to further investigate into the possible superfluidity in 2D co-doped diamond, especially to observe the macroscopic quantum coherence and quantized vortices.

## Conclusion

In conclusion, we have demonstrated the existence of excitonic insulator and possible superfluid phase in B-N co-doped 2D diamond. The resistance rises dramatically by more than three orders and exhibits an obvious phase transition around 220K for one sample. Above $T_C$, the carriers behave as the Arrhenius law predicts, with thermal excitation as the dominant excitation in the system. While temperature cooled down below $T_C$, excitons condense to form BEC due to large exciton binding energy resulting from 2D quantum confinement effect. The variable range hopping model can be applied to describe the transport behavior of the B-N co-doped diamond with a large localization length of 38.9nm, which is in accordance with the characteristic of excitonic insulator. A second phase transition occurs at 61.1K where resistance suddenly drops and even exhibits negative value, and the resistance decreases as an inverse proportional function with biased current at low temperature, possibly indicating the formation of non-equilibrium superfluid state of excitons. Although the existence of the second phase transition in 2D diamond can be drawn herein, further investigations are warranted to explore and confirm the superfluid phase in the system. Our work could open the avenue for searching excitonic insulator in wide bandgap materials, which could further promote the exciton-mediated superconductor or superfluid.

## Acknowledgments

S. Lin thanks the support from the National Natural Science Foundation of China (No. 51202216, 51551203, 61774135 and 62474161). S. Lin thanks for the kind discussion with Andre Geim, Alexy Kavokin and many other colleagues, whose discussions are intriguing.

## Author contributions

S. Lin designed and carried out the experiments, analyzed the data, conceived the study, and wrote the paper. S. Huang, carried out the experiments, analyzed the data, discussed the results and assisted writing the paper. M. Yang participated in the experiments, analyzed the data and discussed the mechanism, X. Chen participated the experiments and discussed the data, H. Bi and K. Xiong participated in the experiments. All authors contributed to the preparation of the manuscript.

## Competing interests

The authors declare that they have no known competing financial interests or personal relationships that could have appeared to influence the work reported in this paper.